# N-body simulation of binary star mass transfer using NVDIA GPUs

**Shaukat Goderya, Tarleton State University, Stephenville, TX, 76402, USA**

**Edward L. Smith, Tarleton State University, Stephenville, TX, 76402, USA**

**Baylor G. Fain, Tarleton State University, Stephenville, TX, 76402, USA**

**Taylor J. Hutyra, Tarleton State University, Stephenville, TX, 76402, USA**

**Mason McCallum, Tarleton State University, Stephenville, TX, 76402, USA**

**Bryant Wyatt, Tarleton State University, Stephenville, TX, 76402, USA**

*Binary star systems are of particular interest to astronomers because they can be used as astrophysical laboratories to study the properties and processes of stars. Between 70% to 90% of the stars in our galaxy are part of a binary star system. Among the many types of binary systems observed, the dynamics of semi-detached and contact systems are the most interesting because they exhibit mass transfer, which changes the composition and life cycle of both stars. The time scales of the mass transfer process are extremely large which makes the process impossible to capture through physical observation. Computer simulations have proved invaluable in refining our understanding of the mass transfer processes. Here we introduce an intuitive, computationally efficient, gravity centered model that simulates the filling of the Roche lobe of an expanding star and its transfer of mass through the first Lagrangian point.*

Binary stars are a system of two stars that orbit a common barycenter. According to statistical estimates, more than 70% of the stars in our galaxy are binary stars and most likely galaxies like our own will also show similar statistics. Our understanding of the theory of stellar evolution of single stars is comprehensive. However, physical models of close binary stars are incomplete and inadequate. There are unsolved problems in the physics of mass transfer, mass loss, and angular momentum evolution that are extremely crucial to our understanding of the origins and evolution of binary stars. Recently, attempts have been made to use binary stars as standard candles to improve the extra-galactic distance scale.

Some binary stars can be observed visually, where the two components can be easily identified and their angular momentum determined. A vast majority of binary stars are not visible. However, if the orbital plane of the system is inclined to the line of sight, the components will display mutual eclipses in their light curve or radial velocity curves. The photometric (light curve) and spectroscopic radial velocity data together allows determination of absolute parameters such as mass, radius, temperature, and other parameters. The most physical model of a binary star is based on the mathematical work carried out by Joseph Lois Lagrange in 1772. The physics of the three-body problem is described by the Lagrangian Equation (1) :

$$L = \frac{1}{2}m_1 v_1^2 + \frac{1}{2}m_2 v_2^2 + \frac{Gm_1}{r_1} + \frac{Gm_2}{r_2} \quad (1)$$

Solving the Lagrangian results in five Lagrangian points, $L_1, L_2, L_3, L_4,$ and $L_5$. The inner Lagrangian points $L_1, L_2, L_3$ are unstable and a small perturbation in forces will cause a point mass to leave the Lagrangian point. The outer Lagrangian points $L_4$ and $L_5$ are stable and hence, a small perturbation in forces will not cause the point mass to leave these Lagrangian points. Edouard Roche in 1873 uses the Lagrangian points to propose Roche equipotential surfaces and pear-shaped Roche lobes. Figure 1 shows the equipotential surfaces and the five Lagrangian points. The equipotential surfaces are given by the Equation (2), where q is the mass ratio $\left(q = \frac{m_2}{m_1}, 0 \leq q \leq 1\right)$. In a semi-detached binary, the Roche lobe makes contact at the inner Lagrangian point $L_1$. The size of these Roche lobes can be calculated from Equation (3).

$$\varphi_n = \frac{2}{(1+q)r_1} + \frac{2}{(1+q)r_2} + \left[y - \frac{q}{(1+q)}\right]^2 \quad (2)$$

$$r_L = \frac{0.49q^{2/3}}{0.69q^{2/3} + \ln(1+q^{1/3})} \quad (3)$$

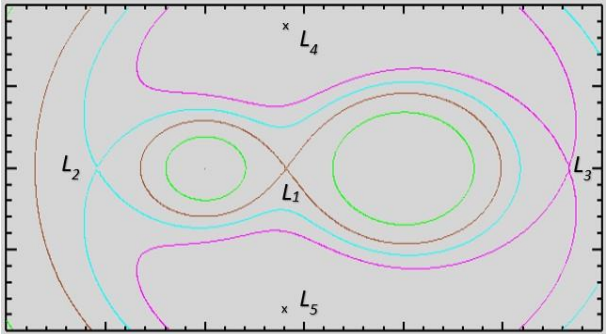

**Figure 1**. Roche potential surfaces and Lagrangian points in a close binary star. Image created with Nightfall Binary Star software.

Equipotential surfaces describe the limits for a particle to orbit two stars and not be captured by one or the other stars. It was Gerard Kuiper in 1941 who recognized the physical significance of the Lagrangian points. Mass transfer can occur between two stars via the inner Lagrangian point $L_1$, and the common envelope around the two stars via the outer Lagrangian points $L_2$ and $L_3$. The process of mass transfer can either be due to Roche-lobe overflow (ROLF) or through wind accretion.

Roche geometry allows us to classify binary stars into three basic morphological groups: detach, semi-detach, and contact. In detach systems, the two stars evolve independently and neither of the Roche lobes are filled up. In semi-detached systems, one of the stars is highly evolved (thus filling its Roche lobe) and is transferring mass to the companion star. Some very interesting cases of semi-detached systems are those in which one of the components is a white dwarf, neutron star, or black hole. In these systems, the compact component will attract matter from the companion that will lead to the formation of an accretion disk around the compact component. Mass transfer in such systems is highly energetic and produce X-ray emission. Figure 2 shows the three morphological groups and their observed light curves. RLOF only happens when the two stars in the system have a regular orbit. Of the observable binary systems, 20 exhibit a more eccentric orbit, making RLOF impossible. Binary systems with eccentric orbits rarely exhibit normal mass transfer through the $L_1$ Lagrangian point. In these cases, the most common points of mass transfer are usually the outer two Lagrangian points, $L_2$ and $L_3$. Even in cases of RLOF, mass transfer can have abnormal behavior under certain conditions. The most common abnormal behavior mimics that of eccentric binary systems in that mass is often lost through the outer two Lagrangian points, sometimes due to an increase in kinetic energy that surpasses the binding energy of the system. This is often seen in Algol-like systems where the mass transfer is considered liberal versus normal. Three main cases of RLOF mass transfer exist: Case A is when the donor star is still in its main sequence and burning hydrogen in its core; Case B when the donor star is starting its first ascent as a red giant on the Hertzsprung-Russell diagram; and Case C when the donor star is on its second ascent as a red giant.

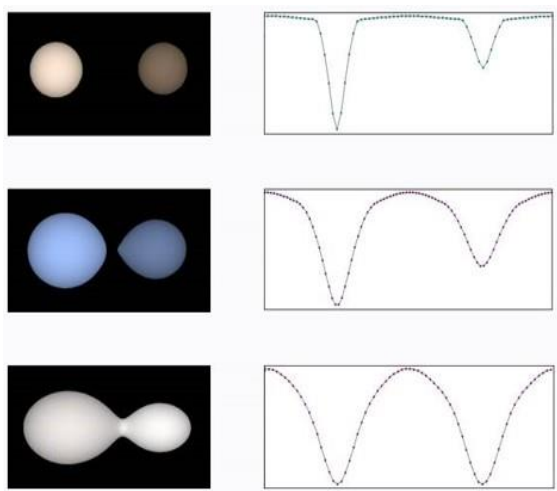

**Figure 2.** Morphological groups of binary stars. Images created with Starlight Pro.

There are many challenges in modeling binary star mass transfer with computers, but the two most important are: the generation of energy in the star via the fusion reaction to account for star expansion, and the different time scales involved in the life cycle and orbital period of the system. One of the most common methods to construct computer models of such nature is to use Smoothed Particle Hydrodynamics (SPH). SPH, which was first developed in the 1970s, works under the assumption that by averaging all of the known values of a smooth function near a point, the value of that point can be found. This process is done using a smoothing kernel, of which there are various forms. While SPH simulations result in fluid, clean simulations, they are computationally intensive and require a high learning curve.

The goal of this paper is to introduce a new, alternative model for binary star mass transfer. Our model uses a Discrete Element Method (DEM) instead of SPH and focuses on the gravitational forces between each particle to power the simulation. Since we use a more brute-force method of simulation, the code implementation for our model compared to SPH models is also much easier without sacrificing the overall accuracy of the model. The end result is meant to be a more intuitive, simpler, and equally valid model that allows a researcher to see in real-time how binary star mass transfer should occur. The work discussed here focuses on the simulation of a semi-detached system in which the primary component is a compact system and the secondary star evolves and transfers mass to the compact system leading to the formation of an accretion disk. Details on the physics of binary stars can be obtained elsewhere.[1,2,3,4]

## METHODOLOGY

The model is based off the work done on lunar forming impact.[5,6] In the model, a star is composed of N quasi-particles, which we refer to as elements (Figure3).[7]

Each element represents a ball of hydrogen gas. The elements each have a set mass and a variable radius. The variable radius is used to simulate a main sequence star growing into a red giant. The properties of the elements are governed solely by the sum total of all the pair-wise element to element interactions. This creates a large N-body system.

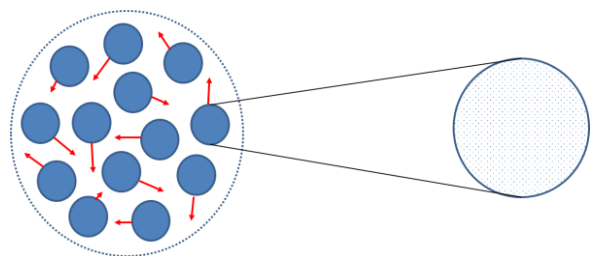

**Figure 3.** Model of a star consisting of N elements

N-body systems are highly parallelizable allowing the compute load to be efficiently distributed across the numerous processors on modern graphics processing units (GPUs).[8] For example, the NVIDIA Geforce RTX 3080 graphics card has a processing power of 29 TeraFlops, for a cost of around $700.

The element-to-element interactions (Figure 4) are described as follows: the elements are always gravitationally attractive. When two elements are in contact with each other and the distance between their centers is decreasing (they are converging), a strong repulsive force proportional to the displaced volume is activated. When two elements that are in contact with each other begin to diverge (the distance between their centers is increasing), the strength of the repulsive force is decreased. This reduction of the repulsive force creates an inelastic collision to simulate an element's transferring some of its translational energy into other forms of energy such as heat. This simplified form of dealing with the thermodynamics of the system greatly reduces the complexity of the simulation code. Figure 4 illustrates the interactions. In Figure 4.A, the force between elements $i$ and $j$ when they are not in contact is given by Equation (4).

**Figure 4.** Force interactions

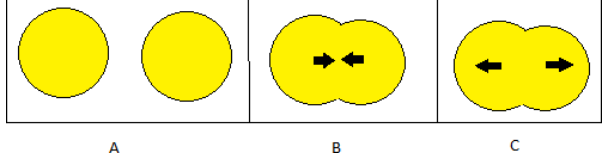

$$F_{i,j} = \frac{Gm_im_j}{(r_{i,j}+\varepsilon)^2} \quad (4)$$

When the elements are in contact and converging as in Figure 4.B, the force between elements *i* and *j* is given by Equation (5).

$$F_{i,j} = \frac{Gm_im_j}{(r_{i,j}+\varepsilon)^2} - \gamma\pi\left(\frac{4}{3}R^3 - R^2r_{i,j} + \frac{r_{i,j}^2}{12}\right) \quad (5)$$

Finally, when the elements are in contact and diverging as in Figure 4.C, the force between elements i and j is given by Equation (6).

$$F_{i,j} = \frac{Gm_im_j}{(r_{i,j}+\varepsilon)^2} - \gamma\beta\pi\left(\frac{4}{3}R^3 - R^2r_{i,j} + \frac{r_{i,j}^2}{12}\right) \quad (6)$$

In these equations, G is the universal gravitational constant, $m_i$ and $m_j$ are the masses of elements *i* and *j*. R is the common radius of the elements; $r_{i,j}$ is the separation of the centers of elements *i* and *j*; $\gamma$ is the strength of the repulsion force; $\beta$ is the reduction fraction in the strength of the repulsion force; and $\varepsilon$ is a small positive number used to remove the singularity in the attractive force.

All elements that create a star are given a common radius. This radius is initially determined by the density of the material that composes the element and the mass of the element. The mass of an element is determined by the mass of the star and the number of elements used to create the star. This technique was used to create a binary system. To simulate a main sequence star evolving into a red giant, the radius of each element of the main sequence star was gradually increased. N-body problems, where n is larger than two, cannot be solved exactly, so a numeric integrator must be used to move the system forward in time. The numeric integrator used in this simulation was the leap-frog formula because it is computationally inexpensive, conservative, and used extensively in computational astrophysics.[9,10]

The main body of the code was written in C and C++. The computationally intensive portions of the simulation were off loaded to the GPU for acceleration. The GPU code was written in compute unified device architecture (CUDA).[11] The graphics used to visualize the simulations were written using OpenGL.[12]

Additionally, GPUs are not only powerful and inexpensive, but fit easily on the PCIe bus of most desktops and workstation. Also, CUDA is easily learned by anyone with basic programming skills. A similar methodology to that described above was adopted to create an isotopically similar Earth-Moon system with correct angular momentum from a giant impact.[5] This paper addressed the concerns of the Philosophical Transactions of the Royal Society A regarding the creation of the Earth-Moon system.

## THEORETICAL MODELS

The evolution of the binary system due to mass transfer can be studied in terms of conservative mass transfer or non-conservative mass transfer. According to theoretical models for non-conservative mass transfer, when the primary component fills its Roche lobe no accretion disk is formed on the secondary component. The secondary components do not evolve, but rather the mass is lost from the system. There are two different cases for this to happen. In case A, the primary component fills its Roche lobe and forms a common envelope (CE) around the two stars leading to either the two components orbiting the common center of mass and the CE getting ejected from the system or the two stars merge with a CE around both of them. In case B, the primary star evolves first and fills its Roche lobe at which point it starts to transfer mass to the secondary star. The secondary star eventually fills its Roche lobe, and begins to transfer mass to the CE envelope.

In conservative mass transfer, the more massive star fills its Roche lobe first at which point a stable mass transfer leads to formation of an accretion disk around the less massive star. The more massive star first becomes a white dwarf, but during this time the less massive star has gained enough mass to become the more massive component, upon which it begins to transfer mass back to the now new less massive component. Astronomers believe that the Algol system is a typical example of this type of mass transfer.

## RESULTS

To test the conservative and non-conservative theoretical models, we considered several different types of binary system configurations and ran the simulations. Figure 5 sequence I-III is an example of a conservative mass transfer. One component is a main sequence and the other is a white dwarf. The sequence of images shows main sequence beginning to transfer mass and forming an accretion disk around the white dwarf. A video of this run can be seen at

Figure 5 sequence IV-VI is an example of a non-conservative case B binary system. The components are initially main-sequence stars and mutually annihilate in the final stage. Figure 5 sequence VII-IX is a conservative mass transfer where one component is a red giant and the other is a main-sequence star. The images show that the red giant begins to transfer mass and forms an accretion disk around the main sequence. This would be an example of an Algol system.

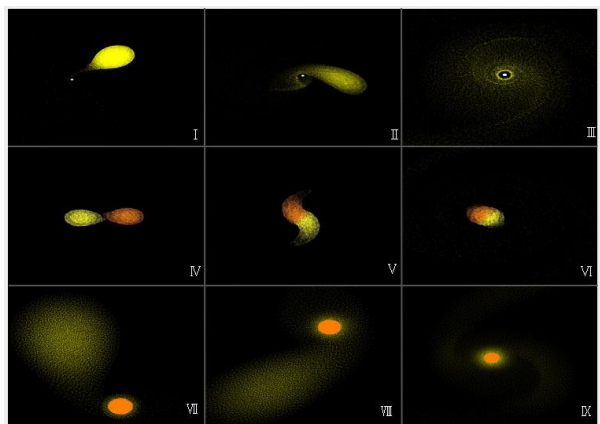

**Figure 5.** Conservative and Non-Conservative simulations

## CONCLUSION

This work shows that the filling of the Roche lobe and subsequent transfer of mass to a companion star in a binary star system, can be simulated with a simple gravitational model. The cost and complexity of creating and running such simulations is small and can be done in most departments. The code can be downloaded from GitHub at: https://github.com/TSUParticleModelingGroup/EvolvingBinaryStarSystemConstantVolume
Links to the videos in figure 5 are below:
https://youtu.be/kDO8LqTGFOs
https://youtu.be/MHRJ8vNUlXw
https://youtu.be/Uh-FYVQ34w8

## ACKNOWLEDGEMENTS

We thank NVIDIA for the donation of hardware; Tarleton State University's high-performance computing lab for time on their computers; Tarleton State University's Division of Research, Innovation and Economic Development, and The Program for Astronomy Education and Research for funding; and special thanks to The Wyatt Fund for supporting student travel.